\documentclass{aa}  
\usepackage{graphicx}
\begin{document}

   \title{Accretion of chemically fractionated material on a wide binary with a blue straggler
          \thanks{Based on observations collected at the European 
                  Southern Observatory, Chile, using FEROS spectrograph
                  (proposal ID: 70.D-0081), on observations made with the Italian Telescopio Nazionale Galileo 
                  (TNG) operated on the island of La Palma by the Fundacion Galileo Galilei of the INAF 
                  (Istituto Nazionale di Astrofisica) at the Spanish Observatorio del Roque de los Muchachos 
                  of the Instituto de Astrofisica de Canarias, and on observations made at McDonald Observatory.}}

   \author{S. Desidera
          \inst{1},
           R.G. Gratton
          \inst{1},
           S. Lucatello
          \inst{1},
           M. Endl
          \inst{2},
           S. Udry
          \inst{3}}

   \authorrunning{S. Desidera et al.}
   \titlerunning{Accretion of chemically fractionated material}

   \offprints{S. Desidera,  \\
              \email{silvano.desidera@oapd.inaf.it} }

   \institute{INAF -- Osservatorio Astronomico di Padova,  
              Vicolo dell' Osservatorio 5, I-35122, Padova, Italy
             \and
             McDonald Observatory, The University of Texas at Austin, Austin, 
             TX 78712, USA
              \and
             Observatoire de Geneve, 51 Ch. des Maillettes, 1290 Sauverny, 
             Switzerland}

 \date{Received  / Accepted }

\abstract
{The components of the wide binary HIP64030=HD 113984 show a large
 (about $0.25$ dex) iron content difference (Desidera et al.~2006 A\&A 454, 581). 
 The positions of the components on the color magnitude diagram suggest
 that the primary is a blue straggler.}
{We studied the abundance difference of several elements besides iron, and we
 searched for stellar and substellar companions around the components
 to unveil the origin of the observed iron difference.}
{A line-by-line  differential abundance analysis for several elements was performed,
  while suitable
 spectral synthesis was performed for C, N, and Li. High precision radial
 velocities obtained with the iodine cell were combined with available
 literature data.}
{The analysis of additional elements shows that the
 abundance difference for the elements studied increases with increasing condensation 
 temperature,
 suggesting that accretion of chemically fractionated 
 material might have occurred in the system.
 Alteration of C and N likely due to CNO processing
 is also observed. We also show that the primary 
 is a spectroscopic binary with a period of 445 days
 and moderate eccentricity. The minimum mass of the companion is $0.17~M_{\odot}$.}
 {Two scenarios were explored to explain the observed abundance pattern. 
  In the first, all  abundance
 anomalies arise on the blue straggler. 
 If this is the case, the dust-gas separation may have been
 occurred in a circumbinary disk around the blue straggler 
 and its expected white dwarf companion, as observed in several 
 RV Tauri and post AGB binaries.
 In the second scenario, accretion of dust-rich material
 occurred on the secondary. This would also explain the anomalous 
 carbon isotopic ratio of the secondary. 
 Such a scenario requires that a substantial amount of mass
 lost by the central binary has been accreted by the wide
 component.
 Further studies to compare the two scenarios are proposed.}
          
   \maketitle


\section{Introduction}
\label{s:intro}

In Desidera et al.~(\cite{chem2}, \cite{chem3}) we performed
high--precision differential abundance analysis of
50 wide visual binaries with similar components.
Only one pair (\object{HIP 64030} = \object{HD 113984} = ADS 8786)
was shown
to have an abundance difference as large as 0.27 dex, with
the secondary being more metal rich. No other pair has an abundance
difference larger than 0.09 dex, and in most cases differences
are lower than 0.03 dex.
This allowed us to conclude that the occurrence of large alterations
of stellar abundances due to the ingestion of metal--rich material
is not a common event, placing limits on the amount of accreted
rocky material similar to estimates of the rocky material
accreted by the Sun during its main sequence lifetime.
The positions of the components of \object{HIP 64030} on the color magnitude
diagram suggest that the primary is a blue straggler.
Therefore, the abundance difference may be somewhat
linked to the peculiar evolutionary history of the system.

Abundance anomalies are expected to occur in blue stragglers formed
through mass transfer events (McCrea \cite{mccrea64} mechanism)
or merging of WUMa binaries, 
while no abundance changes are expected
in case of collisions (Shetrone \& Sandquist \cite{shetrone}). 
For a field blue straggler likely formed through mass transfer,
abundance anomalies of light elements (Li, C, N) are expected
and, in the case of sufficiently massive AGB donor,  
s-elements enhancements are, too.
The appearance of first dredge-up products in the atmosphere of an 
RGB donor would make the star N rich and slightly C poor, altering 
the $^{12}$C/$^{13}$C isotopic ratio at the same time.
Different patterns of carbon and neutron capture enhancements may arise 
depending on the mass and the metallicity of an AGB donor (see, e.g., Norris
et al.~\cite{norris97}).
Iron is not expected to be altered in these processes.
Therefore, other mechanism(s) should play
a  fundamental role in the generation of the 
abundance pattern we observed in \object{HIP 64030}.

The study of several elements besides iron would allow a more complete
picture of the abundance anomalies in this system and
possibly help us to find a full explanation for them.
Such a study is presented in this paper, together with 
a radial velocity (hereafter RV) monitoring of the components 
and a full description of the properties of the system
that can be derived from the available information.

The outline of the paper is as follows: Section~\ref{s:parameters}
presents the basic parameters of the system; in Sect.~\ref{s:bin}
we show that the blue straggler component is a spectroscopic binary
and we derive clues on the orbit of the wide pair; 
Sect.~\ref{s:abu_analysis} presents the procedures adopted
to derive abundance of several elements; 
Sect.~\ref{s:abu}
presents the results of the differential abundance analysis, considering both the
differential analysis between the components and the comparison
with typical stars of similar metallicity and galactic population.
In Sect. \ref{s:discussion} we discuss
 possible scenarios to explain the observed abundance pattern.
Finally, in Sect.~\ref{s:conclusion} we summarize our conclusion
and suggest future developments.

\section{Stellar parameters}
\label{s:parameters}

\begin{figure}
   \includegraphics[width=9cm]{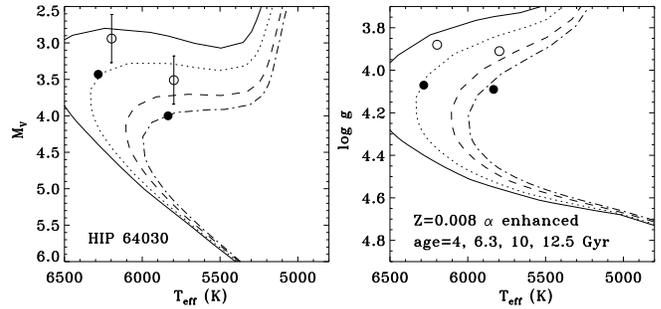}

      \caption{Position of the components of \object{HIP~64030} in the 
               HR diagram (left panel) and in the $T_{\rm eff} - \log g$ 
               diagram (right panel).
               Open circles: results of abundance analysis performed
               adopting the Hipparcos distance; filled circles:
               adopting spectroscopic distances.
               Isochrones by Salasnich et al.~(\cite{salasnich}) with
               Z=0.008 and [$\alpha$/Fe]=0.3, close to the surface
               abundances of the primary, are overplotted. 
               It is not possible to fit both components with a single 
               isochrone.
               The most likely explanation is that the primary is a blue 
               straggler. }
        
         \label{f:cmd}
   \end{figure}

Table \ref{t:star_param} lists the basic
stellar properties of the components of 
the binary system \object{HIP 64030}.
The pair is fairly metal poor.
Our temperatures and gravities imply an anomalous position in the CMD
(Fig.~\ref{f:cmd}).
The primary (brighter component) has properties compatible with
an intermediate age (about 6 Gyr), while the secondary appears
somewhat evolved, indicating an older age (8-12 Gyr).
The details are sensitive to the adopted absolute magnitudes 
(i.e., distance). 
In this paper, we adopt the spectroscopic distance 
(66 pc) instead of that derived from the Hipparcos parallax
(adopted in Desidera et al.~\cite{chem3})
because of the fairly large error of the latter (about 15\%), 
the presence of a short period companion to the primary
discussed in Sect.~\ref{s:bin}, that is not considered
in the derivation of the parallax, and the
better fit of the Balmer lines for the temperature resulting from
the revised distance.
The spectroscopic distance is about 1 $\sigma$ shorter than
the Hipparcos one.
Kinematic parameters and galactic orbit were derived 
as in Barbieri \& Gratton (\cite{bargrat}), considering
the revised distance and averaging radial velocity (Sect.~\ref{s:rv})
and Hipparcos proper motion of the two components.

The enhancement of $\alpha$ elements (see Sect.~\ref{s:abu}) and the kinematics indicate
that the pair is a likely member of the thick disk and thus is very old.
This is fully consistent with the age of the secondary
derived by isochrone fitting, 10-12 Gyr (Fig.~\ref{f:cmd}).
Therefore, the primary appears anomalous, being too bright
for its age. It is thus very likely to be a blue straggler.

Stellar masses were derived using the $\alpha$-enhanced isochrones
by Salasnich et al.~(\cite{salasnich}). Masses of 
$1.04~M_{\odot}$\footnote{This is the mass of the visible component \object{HIP64030Aa},
see Sect.~\ref{s:bin}.}
and $0.96~M_{\odot}$ and ages of about 6 and 10 Gyr for the primary
and secondary, respectively, assuming the surface abundances of
our analysis (Sect.~\ref{s:abu}).
If only the convective zone of the secondary is enriched with heavy elements,
the use of stellar models with the metallicity corresponding to the
surface abundances is not appropriate.
From the stellar models with polluted convective zones (Cody \& Sasselov \cite{cody}),
it results that a larger metal content in the convective zone moves a
star toward lower effective temperatures keeping 
the stellar luminosity nearly constant. 
The mass of the star with a polluted convective zone can be roughly derived
from the isochrone with the chemical composition expected for the inner
parts of the star, assuming the observed absolute magnitude and an
effective temperature warmer than the observed one (by about 150~K for
HIP 64030B, assuming a temperature shift similar to that derived by 
Cody \& Sasselov (\cite{cody}) for a 0.2 dex abundance difference 
for HD 209458).
The mass of HIP 64030B derived in this way results $0.91~M_{\odot}$.
Conversely, if the primary has a stellar atmosphere more metal poor
than the stellar interior, its actual mass would be slightly larger
than $1.04~M_{\odot}$.
In the following, we will assume $M_{Aa}=1.04~M_{\odot}$ and $M_{B}=0.91~M_{\odot}$, 
taking into account that the impact of the different mass assumptions 
on the results of this paper are minor.

\begin{table}[h]
   \caption[]{Stellar properties of the components of \object{HIP 64030}.}
     \label{t:star_param}

       \begin{tabular}{lccc}
         \hline
         \noalign{\smallskip}
         Parameter   &  HIP 64030~A &  HIP 64030~B  & Ref. \\
         \noalign{\smallskip}
         \hline
         \noalign{\smallskip}

$\mu_{\alpha}$ (mas/yr)  & -101.92 $\pm$ 1.66 &  -97.60 $\pm$ 2.36  & 1   \\
$\mu_{\delta}$ (mas/yr)  & -105.92 $\pm$ 1.40 & -102.24 $\pm$ 2.20  & 1   \\
RV     (km/s)            &  -90.1 $\pm$ 0.2 & -89.6 $\pm$ 0.2 & 2 \\ 
$\pi$  (mas)             & \multicolumn{2}{c}{13.24 $\pm$ 1.99}   & 1   \\
$d_{spec}$    (pc)       & \multicolumn{2}{c}{66 $\pm$ 10 }  & 2 \\  
$U$   (km/s)             & \multicolumn{2}{c}{-38.3 $\pm$ 2.0 }  & 2 \\
$V$   (km/s)             & \multicolumn{2}{c}{-11.1 $\pm$ 6.5 }  & 2 \\
$W$   (km/s)             & \multicolumn{2}{c}{-91.9 $\pm$ 2.0 }  & 2 \\
$R_{min}$   (kpc)        & \multicolumn{2}{c}{ 7.84 $\pm$ 0.16 } & 2 \\
$R_{max}$   (kpc)        & \multicolumn{2}{c}{ 9.77 $\pm$ 0.34 } & 2 \\
$e$                      & \multicolumn{2}{c}{ 0.11   }  & 2 \\
$z_{max}$   (kpc)        & \multicolumn{2}{c}{ 1.98 $\pm$ 0.07 }  & 2 \\

 & &  &   \\
V                        & 7.528 $\pm$ 0.004  &  8.095 $\pm$ 0.007 & 3 \\
$(B-V)$                  & 0.439 $\pm$ 0.013  &  0.539 $\pm$ 0.024 & 3 \\
$H_{p}$                  & 7.639 $\pm$ 0.005  &  8.219 $\pm$ 0.008 & 1  \\
$H_{p}$ scatter          & \multicolumn{2}{c}{0.010$^{\mathrm{a}}$} & 1 \\
J                        & 6.596 $\pm$ 0.018 & 6.961 $\pm$ 0.020 & 4 \\
H                        & 6.409 $\pm$ 0.044 & 6.703 $\pm$ 0.033 & 4 \\
K                        & 6.325 $\pm$ 0.017 & 6.610 $\pm$ 0.023 & 4 \\
$b-y$                    & \multicolumn{2}{c}{0.332$^{\mathrm{a}}$  $\pm$ 0.004} & 5 \\
$m_{1}$                  & \multicolumn{2}{c}{0.126$^{\mathrm{a}}$  $\pm$ 0.005} & 5 \\
$c_{1}$                  & \multicolumn{2}{c}{0.398$^{\mathrm{a}}$  $\pm$ 0.006} & 5 \\
 & & &  \\
ST                       & F5V  & F9V  &  6  \\
$M_{V}$                  & 3.47$\pm$0.25  &    4.04 $\pm$0.25 & 2 \\  

$T_{eff}$ (K)            & 6283$\pm$100  &  5834$\pm$100  &  2 \\
$\Delta T_{eff}$ (K)     & \multicolumn{2}{c}{   451$\pm$20   } &  2 \\

$\log g$                 &      4.07$\pm$0.15   &  4.09$\pm$0.15  &  2 \\
 & & &    \\
$\log R^{'}_{HK}$        & -5.06   &  -5.12   &  7 \\
$ v \sin i $ (km/s)      &    4.1     &        2.5  &  7 \\
$L_{X}$ (erg/s)          & \multicolumn{2}{c}{  $<1.0~10^{29~a}$} & 2 \\

 & & &   \\
${\rm [Fe/H]}$           &       -0.57$\pm$0.10      &       -0.32$\pm$0.10      &  2 \\
$\Delta {\rm [Fe/H]}$     & \multicolumn{2}{c}{ -0.251$\pm$0.020  } &  2 \\
 & & &   \\
${\rm Mass} (M_{\odot})$ &  1.04$\pm$0.05$^b$   &  0.91$\pm$0.05   &  2 \\
                         &  $>1.21^c$           &                  &  2 \\

Age  (Gyr)  (isoc.)  &  $\sim 6$   & 8-12  &  2  \\
Age  (Gyr)  ($R_{HK}$)   & 7       & 8     &  7  \\

         \noalign{\smallskip}
         \hline
      \end{tabular}

References: 1 Hipparcos (ESA \cite{hipparcos});
            2 This paper;
            3 Desidera et al.~(\cite{chem3});
            4 2MASS (Cutri et al.~\cite{2mass});
            5 Olsen (\cite{olsen});
            6 Abt (\cite{abt81});
            7 Desidera et al.~(\cite{binferos}).

\begin{list}{}{}
\item[$^{\mathrm{a}}$] A+B
\item[$^{\mathrm{b}}$] Aa
\item[$^{\mathrm{c}}$] Aa+Ab
\end{list}
\end{table}

Blue stragglers and ultra--lithium depleted stars  often show moderate
rotational velocity (Fuhrmann \& Bernkopf \cite{fuhrmann99}; 
Ryan et al.~\cite{ryan02}). This is thought to
arise because of the acquisition of angular momentum during the
mass transfer process.
In \object{HIP 64030}, the BS component shows a slightly enhanced 
$v \sin i$ (4.1 km/s) with respect to the secondary, a slow rotator 
as expected for a very old star ($v \sin i=$ 2.5 km/s)\footnote{These 
values are calculated for the B-V
color obtained by Tycho. Adopting the colors corresponding
to the spectroscopic temperatures increases the rotational
velocities by just 0.2 km/s because of the smaller macroturbulent
velocity.}. 
The low chromospheric emission, the lack
of X--ray detection by ROSAT (Voges et al.~\cite{rosat}), and the small photometric variability
from Hipparcos are further indications of a fairly low
rotation rate.            

\section{Binarity}
\label{s:bin}

The binary fraction among field blue stragglers is very high
(Preston \& Sneden \cite{preston00}; Carney et al.~\cite{carney05}).
The lack of double--lined spectroscopic binaries and the distribution of 
the mass function suggest that most of the companion are
white dwarfs with masses of about $0.55~M_{\odot}$.
The period and eccentricity distributions are also different with
respect to a comparison sample of normal main sequence stars
(Duquennoy \& Mayor \cite{duquennoy}). Most of the blue stragglers
have periods in the range 100 to 3000 days and  eccentricities
smaller than those of field binaries. The low eccentricity can be explained
with the occurrence of angular momentum dissipation as part of the mass transfer 
process (e.g., McClure \cite{mcclure97}).
All these properties are expected if these objects were formed 
predominately via the McCrea (\cite{mccrea64}) scenario.
If \object{HIP 64030A} is indeed a blue straggler, as we are claiming, 
we then expect it has a white dwarf companion with short orbital period
and low eccentricity.
In this section we consider the evidence in favor of this
hypothesis that results from radial velocities and astrometric
measurement, and the clues on the orbit of the wide pair.

\subsection{Radial velocity}
\label{s:rv}

The available RV measurements taken with traditional techniques (errors
larger than 0.2 km/s), including previously unpublished CORAVEL 
data, suggest the occurrence of low-amplitude radial velocity
variations for \object{HIP 64030A} (Table \ref{t:rv}).
Conclusive evidence of the radial velocity variability comes from 
ten spectra taken with SARG, the high resolution spectrograph of the TNG
(Gratton et al.~\cite{sarg}), and at the Harlan J.~Smith 2.7 m
telescope at McDonald Observatory. 
These spectra were acquired with the iodine cell and analyzed as 
those for the on-going planet search programs  using these instruments 
(Desidera et al.~\cite{desidera06}; Endl et al.~\cite{austral}; 
Wittenmyer et al.~\cite{witt06}). Errors in differential radial
velocities are about 10 m/s and 15 m/s for SARG and McDonald, respectively, 
slightly larger than is typical for stars of similar magnitude observed with 
these instruments because of the 
shallowness of the spectral lines of this F5 metal-poor star.
The differential radial velocities measured with the iodine cell technique
were placed on an absolute scale to allow a combination of the whole data set
by cross-correlation using the spectral orders between 4700 and 4970 \AA \, that are 
free of iodine lines, further adjusting the zero-point of the McDonald spectra 
using telluric lines.

\begin{figure}
   \includegraphics[width=9cm]{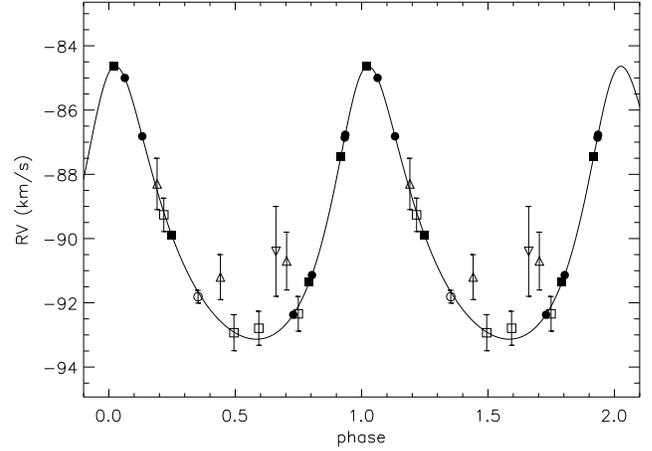}
      \caption{The radial velocities of HIP 64030 A phased to the preliminary
              orbital  period (445 days). The best--fit orbital solution is overplotted.
              Radial velocities from different sources are represented by different
              symbols: filled circles: SARG (this paper); filled squares: McDonald 
              (this paper); open circle: FEROS (Desidera et al.~\cite{binferos}); empty squares:
              CORAVEL (this paper); empty upper triangles: Struve \& Zebergs (\cite{struve59}); 
              upside down triangle: Andersen \& Nordstrom (\cite{andersen85}).}
        
         \label{f:phase}
   \end{figure}

The highest peak of the Scargle-Lomb periodogram of the radial velocities 
is at 445 days.
A Keplerian fitting yields the orbital parameters listed in Table~\ref{t:orbit}.
The radial velocities phased to this  orbital period are shown in
Fig.~\ref{f:phase}. A Keplerian fitting for other possible periods resulting from
the periodogram clearly indicates that the 445 days period is the real one.

The minimum mass of the companion is about $0.17~M_{\odot}$.
This is compatible with the presence of a white dwarf companion,
as expected for blue stragglers formed through mass transfer from 
an RGB companion.
The orbital period and eccentricity are also normal for a field
blue straggler (Preston \& Sneden \cite{preston00}).
The secondary instead shows  a constant RV, fully compatible with the
observational errors, both on decade timescales with errors of
about 1 km/s and on the two-year timespan of SARG observations with
errors of about 10 m/s.
Hereafter we refer to \object{HIP 64030Aa} as the current primary star
(the blue straggler), to \object{HIP 64030Ab} as the unseen spectroscopic
companion (which we assume is a low mass white dwarf), and to  \object{HIP 64030B}
as the wide visual companion.

\begin{table}
   \caption[]{Radial velocity of the components of \object{HIP 64030}.}
     \label{t:rv}
      
       \begin{tabular}{cccl}
         \hline
         \noalign{\smallskip}
         Epoch      &  RV (A) &  RV(B)  & Ref. \\
         \noalign{\smallskip}
         \hline
         \noalign{\smallskip}

$<34379$  & $-92.2\pm1.9$            & $-87.0\pm1.5$             & 1 \\  
34817.50  & $-88.3\pm0.8^a$          & $-89.2\pm0.9^a$             & 2 \\ 
35856.50  &                          & $-90.0\pm0.6^a$             & 2 \\ 
35937.50  & $-90.7\pm0.9^a$          &                           & 2 \\  
36266.50  & $-91.2\pm0.7^a$          & $-90.4\pm0.8^a$             & 2 \\ 
43912.695 & $-92.53\pm0.53$          & $-89.51\pm0.48$           & 3 \\
44314.588 & $-92.67\pm0.56$          & $-90.42\pm0.47$           & 3 \\
44388.62  & $-90.4 \pm1.4 $          & $-89.1\pm1.4 $            & 4 \\
44636.696 & $-89.00\pm0.52$          & $-89.13\pm0.39$           & 3 \\
49777.535 & $-92.08\pm0.54$          & $-89.62\pm0.48$           & 3 \\
52721.16  & $-91.8 \pm0.2 $          & $-89.6 \pm0.2 $           & 5 \\
53483.61  & $-84.999\pm0.008^b$      & $-89.625\pm0.004^b$       & 6 \\
53514.43  & $-86.817\pm0.006^b$      & $-89.622\pm0.004^b$       & 6 \\
53566.70  & $-89.746\pm0.015^b$      &                           & 7 \\
53780.78  & $-92.372\pm0.012^b$      &                           & 6 \\  
53808.88  & $-91.205\pm0.018^b$      &                           & 7 \\
53813.70  & $-91.138\pm0.009^b$      & $-89.629\pm0.009^b$       & 6 \\
53864.82  & $-87.288\pm0.017^b$      &                           & 7 \\
53871.52  & $-86.857\pm0.006^b$      & $-89.614\pm0.004^b$       & 6 \\
53872.51  & $-86.769\pm0.014^b$      & $-89.607\pm0.004^b$       & 6 \\
53910.72  & $-84.485\pm0.014^b$      &                           & 7 \\

        \noalign{\smallskip}
         \hline
      \end{tabular}

References:  1 Wilson  (\cite{wilson});
             2 Struve \& Zebergs (\cite{struve59});
             3 CORAVEL (this paper);
             4 Andersen \& Nordstrom (\cite{andersen85});
             5 FEROS (Desidera et al.~\cite{binferos});
             6 SARG (this paper);
             7 McDonald (this paper).

\begin{list}{}{}
\item[$^{\mathrm{a}}$] Internal errors are quoted in the original
reference. Comparison with Nordstrom et al.~(\cite{nordstrom}) for
stars in common that are not spectroscopic binaries indicates that
the RV zero point is the same and that a further 0.5 km/s  error 
should be present. Such a error was added in quadrature to the internal errors.

\item[$^{\mathrm{b}}$] Internal errors of differential radial velocities
obtained using iodine lines. Errors on absolute radial velocities
are about 0.2-0.3 km/s.

\end{list}
\end{table}

\begin{table}[h]
   \caption[]{Preliminary orbital solution for  \object{HIP 64030A}.}
     \label{t:orbit}

       \begin{tabular}{lc}
         \hline
         \noalign{\smallskip}
         Parameter   &  Value \\
         \noalign{\smallskip}
         \hline
         \noalign{\smallskip}

Period (d)                         & $445.8\pm0.8$\\
RV Semi-amplitude (km/s)           & $4.25\pm0.08$\\
Eccentricity                       & $0.29\pm0.01$\\
Longitude of periastron (deg)      & $342.5\pm2.3$ \\
Periastron passage (JD-2400000)    & $ 43648\pm18$ \\
$\gamma$ (km/s)                    & $-90.06\pm0.07$\\
RV offset (McDonald-SARG) (km/s)   & $-0.158\pm0.010$  \\
RV offset (CORAVEL-SARG)  (km/s)   & $-0.26\pm0.30$  \\
Mass function  ($M_{\odot}$)       & $0.00311\pm0.00018$ \\
                                   & \\
$a$ (AU)                           & $1.215\pm0.002$ \\ 
$m \sin i~~~(M_{\odot})$           & $0.165\pm0.004$ \\ 

         \noalign{\smallskip}
         \hline
      \end{tabular}
\end{table}

\subsection{Astrometry}
\label{s:astrometry}

\begin{figure}
   \includegraphics[width=9cm]{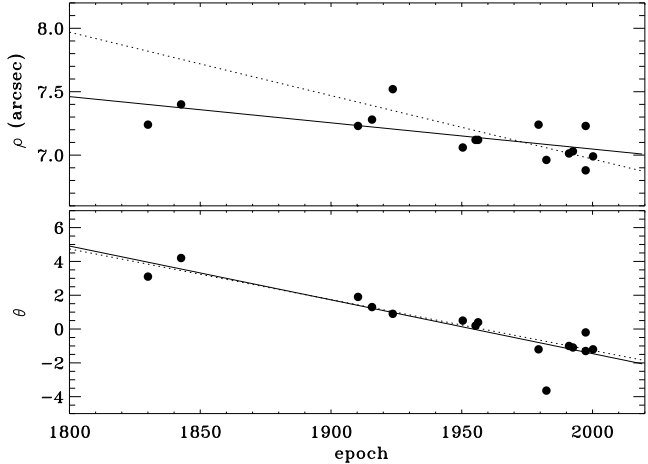}
      \caption{Relative motion of \object{HIP 64030 A and B}. The continuous line is
               the linear fitting to the data (without weighting).
               The dashed line is the relative motion measured by Hipparcos
               (extrapolated to a timescale much larger than the mission
               lifetime).}
        
         \label{f:rhotheta}
   \end{figure}

Relative astrometry of the components allows us to place constraints on
the orbital motion of the wide pair. Furthermore, it provides additional
evidence of the binarity of the primary.
Hipparcos detected relative motion between \object{HIP 64030 A and B}
($d \rho / dt = -0.005 $ arcsec/yr and $d \theta / dt = -0.030 $ deg/yr).
The literature position measurements (Table \ref{t:astrometry})
are consistent with Hipparcos
results concerning position angle, while the Hipparcos separation  
gradient is too steep to fit the old historical data reported
by Aitken (\cite{aitken}), as shown in Fig.~\ref{f:rhotheta}.

\begin{table}
   \caption[]{Relative position of the components of \object{HIP 64030}.}
     \label{t:astrometry}
      
       \begin{tabular}{cccc}
         \hline
         \noalign{\smallskip}
         Epoch      &  $\rho$ & $\theta$ & Ref. \\
         \noalign{\smallskip}
         \hline
         \noalign{\smallskip}

1830.01  & 7.24  &          3.1       &        1 \\
1842.68  & 7.40  &          4.2       &        1 \\
1910.33  & 7.23  &          1.9       &        1 \\
1915.68  & 7.28  &          1.3       &        1 \\
1923.62  & 7.52  &          0.9       &        1 \\
1950.38  & 7.06  &          0.5       &        2 \\
1955.27  & 7.12  &          0.2       &        3 \\
1956.24  & 7.12  &          0.4       &        3 \\
1979.35  & 7.24  &        358.8       &        4 \\
1982.35  & 6.96  &        356.4       &        5 \\
1991.25  & 7.013 &        359.0       &        6 \\
1992.50  & 7.03  &        358.91      &        7 \\
1997.36  & 6.88  &        358.7       &        8 \\   
1997.37  & 7.23  &        359.8       &        8 \\   
2000.19  & 6.99  &        358.8       &        9 \\

         \noalign{\smallskip}
         \hline
      \end{tabular}

References:  1 Aitken  (\cite{aitken});
             2 Muller (\cite{muller});
             3 Gasteyer \& Eichhorn, (\cite{gasteyer});
             4 Heintz (\cite{heintz80});
             5 Pannunzio \& Morbidelli (\cite{pannunzio});
             6 Hipparcos (ESA \cite{hipparcos});
             7 Van Dessel \& Sinachopoulos (\cite{vandessel93});
             8 Alzner (\cite{alzner});                                 
             9 2MASS  (Cutri et al.~\cite{2mass}).

\end{table}

Part of the relative motion between A and B measured by Hipparcos
is probably due to the contribution of \object{HIP 64030Ab}.
The impact of this on the system parallax as derived by Hipparcos is
difficult to assess. In some cases of unrecognized multiplicity,
large parallax errors were discovered (see, e.g., 
S\"oderhjelm \cite{soderhjelm99}).

\begin{figure}
   \includegraphics[width=9cm]{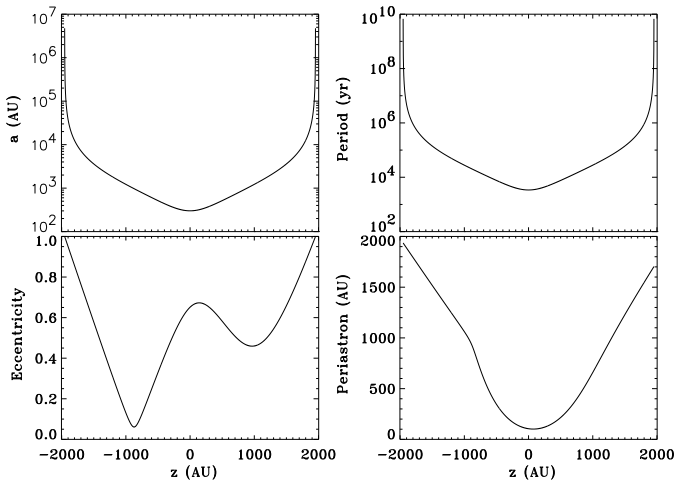}
      \caption{Possible orbital parameters of the wide binary as a function of the
               (unknown) separation along the line of sight.}
         \label{f:orbit}
   \end{figure}

The long--term relative motion of the wide binary coupled with
the RV difference can be used to constrain the orbit
(Hauser \& Marcy \cite{hauser}; Desidera et al.~\cite{hd219542}).
Assuming a $0.4~M_{\odot}$ companion for the spectroscopic companion of
the blue straggler, a broad
range of orbital parameters is compatible with the data (Fig.~\ref{f:orbit}).
Orbits with periastron down to about 100 AU are possible.

\section{Abundance analysis}
\label{s:abu_analysis}

\subsection{Atmospheric parameters and iron abundance}

The binary system \object{HIP 64030} shows a large abundance difference
with the secondary
being significantly more metal rich (Desidera et al.~\cite{chem3}).
The analysis performed adopting the spectroscopic distance
yields the following atmospheric parameters:
for the primary $T_{\rm eff}=6283$ K, $\log g=4.07$, [A/H]=-0.57,
microturbulent velocity = 1.71 km/s; for the secondary $T_{\rm eff}=5834$ K, 
$\log g=4.09$, [A/H]=-0.32,
microturbulent velocity = 1.16 km/s.
The effective temperatures are warmer (by about 90 K) and gravities higher 
(by about 0.15 dex) than
the analysis presented in Desidera et al.~(\cite{chem3}) because of
the fainter absolute magnitude adopted in the analysis.
The stellar masses derived in Sect.~\ref{s:parameters} were used to derive
stellar gravities.
The iron abundance difference result $-0.251$ dex.
Performing the analysis with different assumptions (different absolute
magnitude, optimization of microturbulence instead of using
a $\xi-T_{\rm eff}$ relation, using B as reference, exclusion of 
vanadium lines to derive effective temperature) indicates that
errors on this difference should not exceed 0.04 dex.
The difference of effective temperature based on different techniques
agree within errors.
An error in temperature difference as large as 300 K would be required
to eliminate the abundance difference.

The possibility that blending causes a spourious abundance difference
also does not seem viable.
Even if the primary is composed of a pair of equal components (i.e.,
equal spectrum with magnitude difference of 0.75 mag), the abundance
difference and the anomalous position on the HR diagram could not be  eliminated. 
A blend of the primary with a red star would make the age
difference between the primary and the secondary even worse .
In any case, there is no indication of contributions by the faint
companion HIP 64040Ab on our spectra
(asymmetries of line profiles, trends with wavelength of the observed
abundance difference).

\subsection{Analysis of other elements}
\label{s:other_elements}

The wide spectral coverage of the FEROS spectrograph allows
the inclusion of spectral lines of several elements
besides iron. The measurement of elements other than iron
is useful to better constrain the properties of the stars
and the origin of the observed iron difference.

\subsubsection{Analysis based on EW measurement}

Standard analysis based on EW measurement was performed
for several elements.
Hyperfine structure is taken into account for Sc, V, Mn, and Ba
(see Gratton et al.~\cite{gratton03}).
Differential line-by-line analysis was performed as for iron
(Desidera et al.~\cite{chem3}).

\subsubsection{Spectral synthesis of light elements}

\begin{figure}
    \includegraphics[width=9cm,height=12cm]{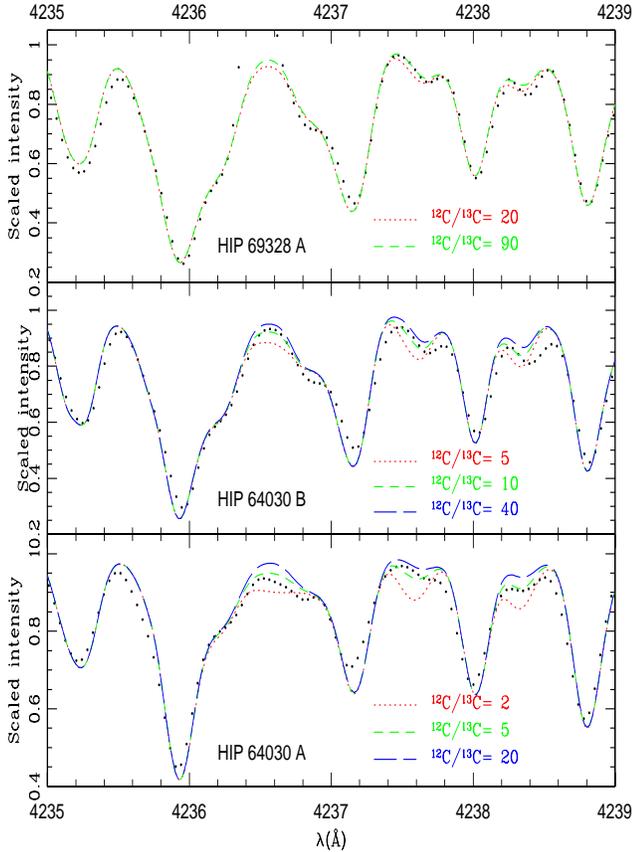}
       \caption{A portion of the spectra used to derive the C isotopic ratio using spectral synthesis.
                Upper panel: the comparison star HIP 69328A, central panel: HIP 64030B, lower panel:
                HIP 64030 A.}
          \label{f:cisob}
    \end{figure}

 From the synthesis of different spectral ranges,
 we obtained measurements for the abundances of Li, C, N, and O,
  as well as  estimates on the C isotopic ratio.
 The line lists for the syntheses have been built as described in
 Lucatello et al.~(\cite{lucatello03}), starting from the Kurucz line lists
 applying small wavelength and $\log gf$ adjustments to reproduce the solar
 spectrum.
 C abundance and C isotopic ratios were measured from several
 CH lines in the G-band around 4300\,\AA. An example of the features fitted
 is given in Fig.~\ref{f:cisob}.
 The C abundance results [C/Fe]=$+0.20\pm0.10$ and [C/Fe]=$-0.10\pm0.10$ for
 the primary and the secondary, respectively.
 The C isotopic ratio results  $^{12}$C/$^{13}$C$=5-10 \pm 5$
 for the primary and  $^{12}$C/$^{13}$C$=30-40 \pm 10$ for the secondary.
 While the low isotopic ratio is expected for a blue straggler
 (see Sect.~\ref{s:abu}), it is unexpected for the secondary.
 We checked our synthesis procedure using a spectrum taken on
 the same run of the star \object{HIP 69328 A}, which has atmospheric
 parameters fairly similar to  \object{HIP 64030 B} 
 (Desidera et al.~\cite{chem3}).
 A $^{12}$C/$^{13}$C of 30-40 is clearly excluded for this star
 (Fig.~\ref{f:cisob}, upper panel).

\begin{figure}
    \includegraphics[width=9cm]{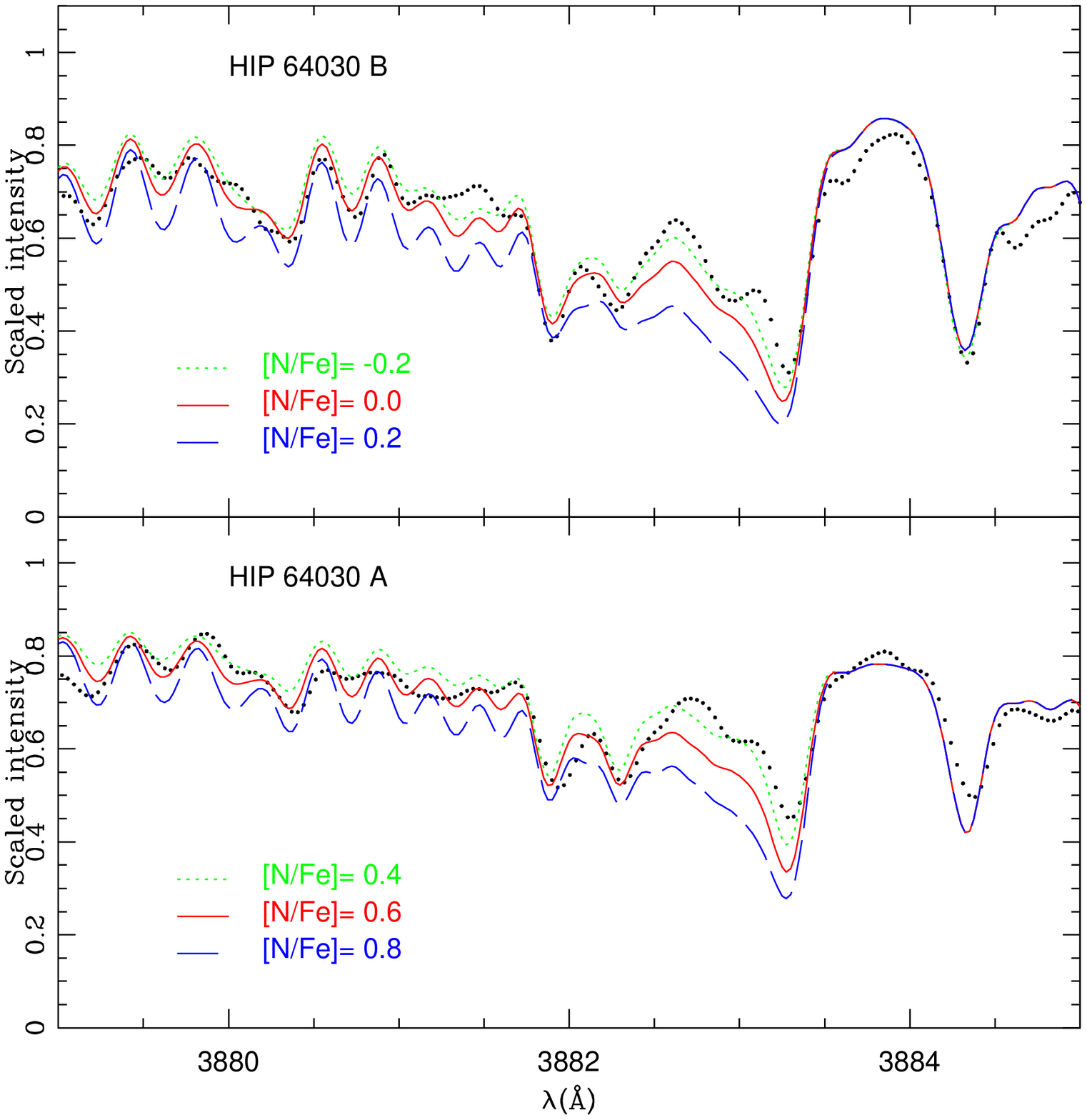}
       \caption{CN spectral synthesis.
                Upper panel: HIP 64030B, lower panel:
                HIP 64030 A.}
          \label{f:cn}
    \end{figure}

 The UV CN system at 3860 \AA~was used to derive N abundance
 (Fig.~\ref{f:cn}). The N abundance results
 [N/Fe]=$+0.50\pm0.15$ and [N/Fe]=$-0.10\pm0.15$
 for the primary and secondary, respectively.
 The O abundance was measured by synthesizing the
 $\lambda 6300.31$~\AA~forbidden line. To
 account for the contribution of the telluric lines to the spectra,
 we subtracted a properly scaled early--type, fast rotating star spectrum
 taken during the same observing run.
 The O abundance results [O/Fe]$\leq +0.40$ and [O/Fe]$=+0.20\pm0.15$
 for the primary and secondary, respectively. It should be noted
that the subtraction of the telluric lines prevents the accurate
determination of abundance for O we would expect from spectra of the
quality of the present data.
 Li was measured from the 6707\,\AA~resonance line; however,
 in the case of HIP64030A we could only derive an upper limit to it
(A(Li)$<1.8$). For the secondary we found A(Li)$=2.7\pm0.1$.

  \begin{figure}
    \includegraphics[width=9cm]{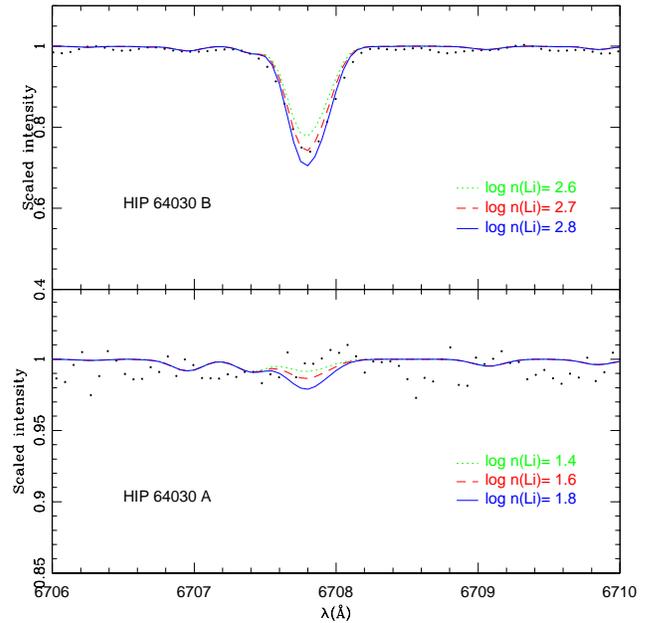}
       \caption{Li spectral synthesis. Upper panel: HIP 64030B, lower panel:
                HIP 64030 A.}
          \label{f:li}
    \end{figure}

\subsection{Sensitivity of abundance differences to atmospheric parameters} 

Table \ref{t:sens_abu} shows the sensitivity of the abundance difference of 
the elements included in our analysis to variations of atmospheric parameters.
The values were obtained by changing the atmospheric parameters of the secondary
one by one.
Different species have quite different sensitivities, indicating that it is not
possible to erase the observed abundance difference for all the species
changing the atmospheric parameters.
      
\begin{table*}
   \caption[]{Sensitivity of abundance differences to atmospheric parameters.}
     \label{t:sens_abu}
      
{\scriptsize

       \begin{tabular}{ccccccccc}
         \hline
         \noalign{\smallskip}
         Element  & $\Delta$ (A-B) &  $\Delta$ (A-B) &   $\Delta$ (A-B) & $\Delta$ (A-B) &$\Delta$ (A-B) &$\Delta$ (A-B) &$\Delta$ (A-B) &$\Delta$ (A-B) \\
                  & $\Delta T_{\rm eff}$ & $\Delta T_{\rm eff}$ & $\Delta \log g$  & $\Delta \log g $ & $\Delta$ [A/H] &  $\Delta $[A/H]  & $\Delta \xi $ & $\Delta \xi $ \\
                  & $+100$~K  & $-100$~K  & $+0.10$~dex & $ -0.10$~dex & $+0.10$~dex & $ -0.10$~dex & $ +0.20$~km/s & $ -0.20$~km/s \\

         \noalign{\smallskip}
         \hline
         \noalign{\smallskip}

Fe I   & -0.0810 &  0.0654 & -0.0103 & -0.0101 & -0.0037 &  0.0010 &  0.0278 & -0.0553 \\
Fe II  &  0.0265 & -0.0152 & -0.0328 &  0.0445 & -0.0145 &  0.0271 &  0.0574 & -0.0383 \\
C I    &  0.0544 & -0.0593 & -0.0346 &  0.0324 &  0.0044 & -0.0060 &  0.0010 & -0.0020 \\
Na I   & -0.0485 &  0.0530 &  0.0065 & -0.0060 & -0.0015 &  0.0035 &  0.0110 & -0.0105 \\
Mg I   & -0.0400 &  0.0434 &  0.0044 & -0.0026 & -0.0006 &  0.0004 &  0.0084 & -0.0066 \\
Al I   & -0.0420 &  0.0440 &  0.0005 & -0.0020 & -0.0010 & -0.0010 &  0.0035 & -0.0055 \\
Si I   & -0.0296 &  0.0294 &  0.0005 &  0.0025 & -0.0056 &  0.0075 &  0.0117 & -0.0065 \\
Si II  &  0.0735 & -0.0795 & -0.0265 &  0.0285 & -0.0045 &  0.0080 &  0.0135 & -0.0075 \\
S I    &  0.0780 & -0.0543 & -0.0357 &  0.0333 &  0.0013 & -0.0023 &  0.0043 & -0.0053 \\
Ca I   & -0.0727 &  0.0711 &  0.0173 & -0.0153 & -0.0027 &  0.0034 &  0.0397 & -0.0440 \\
Ca II  &  0.0527 & -0.0523 & -0.0247 &  0.0490 &  0.0007 & -0.0037 & -0.0023 & -0.0003 \\
Sc II  & -0.0004 &  0.0102 & -0.0335 &  0.0472 & -0.0190 &  0.0333 &  0.0531 & -0.0143 \\
Ti I   & -0.1059 &  0.1051 &  0.0053 & -0.0118 &  0.0006 &  0.0001 &  0.0434 & -0.0477 \\
Ti II  &  0.0115 &  0.0073 & -0.0332 &  0.0478 & -0.0163 &  0.0299 &  0.0493 & -0.0474 \\
V I    & -0.1670 &  0.1065 &  0.0011 & -0.0006 & -0.0002 &  0.0011 &  0.0124 & -0.0124 \\
V II   &  0.0135 & -0.0035 & -0.0335 &  0.0460 & -0.0115 &  0.0245 &  0.0215 & -0.0080 \\
Cr I   & -0.0850 &  0.0845 &  0.0050 & -0.0039 &  0.0001 &  0.0003 &  0.0442 & -0.0532 \\
Cr II  &  0.0247 & -0.0275 & -0.0334 &  0.0429 & -0.0126 &  0.0234 &  0.0437 & -0.0326 \\
Co I   & -0.0973 &  0.1269 & -0.0017 &  0.0030 & -0.0383 &  0.0040 & -0.0207 & -0.0787 \\
Ni I   & -0.0542 &  0.0621 &  0.0006 &  0.0026 & -0.0026 &  0.0051 &  0.0205 & -0.0157 \\
Cu I   & -0.0140 &  0.0530 & -0.0030 &  0.0427 & -0.0045 &  0.0457 &  0.0135 &  0.0247 \\
Zn I   &  0.0756 &  0.0260 & -0.0074 &  0.0146 & -0.0137 &  0.0216 &  0.0600 &  0.0550 \\
Sr I   & -0.0077 &  0.0230 & -0.0327 &  0.0490 & -0.0190 &  0.0363 & -0.0295 & -0.0790 \\
Y II   & -0.1010 &  0.1080 &  0.0020 &  0.0000 & -0.0010 &  0.0020 &  0.0110 & -0.0110 \\
Zr II  & -0.0020 &  0.0120 & -0.0350 &  0.0490 & -0.0200 &  0.0330 &  0.0120 &  0.0020 \\
Ba II  & -0.0260 &  0.0300 & -0.0320 &  0.0430 & -0.0200 &  0.0300 &  0.1800 & -0.1950 \\
La II  & -0.0180 &  0.0320 & -0.0340 &  0.0510 & -0.0210 &  0.0400 &  0.0190 & -0.0030 \\
Ce II  & -0.0180 &  0.0320 & -0.0330 &  0.0496 & -0.0200 &  0.0390 &  0.0226 & -0.0100 \\
Nd II  & -0.0206 &  0.0354 & -0.0323 &  0.0500 & -0.0210 &  0.0407 &  0.0197 & -0.0030 \\

         \noalign{\smallskip}
         \hline
      \end{tabular}

}

\end{table*}

\section{Abundances of HIP 64030 A and B}
\label{s:abu}

\subsection{Relative abundances with respect to iron}

As we will discuss in Sect.~\ref{s:discussion}, it is not easy to
understand which of the components can be considered the
unpolluted one.
The relative abundance with respect to iron may help
for this and for a better classification of galactic population
of our target.
Table \ref{t:diffabu} reports the results.
Both components show enhancements of $\alpha$ elements (Mg, Si, Ca).
This supports the kinematic association to the thick disk
population (and then the very old age for the system and the 
blue straggler status for the primary).

Nitrogen is strongly enhanced and the $^{12}$C/$^{13}$C isotopic ratio 
is very low in the primary.
We think these are the signatures of mass transfer
from material processed through the CNO cycle within the companion, and 
dredged up on the
surface during RGB phase.
However, CNO processing should also produce  some C depletion, which 
is not observed.
The low lithium content of the primary
is typical of  blue stragglers (Ryan et al.~\cite{ryan01}; 
Carney et al.~\cite{carney05}), probably because
complete lithium depletion occurred in the atmosphere
of the RGB/AGB donor.

\begin{figure}
   \includegraphics[width=9cm]{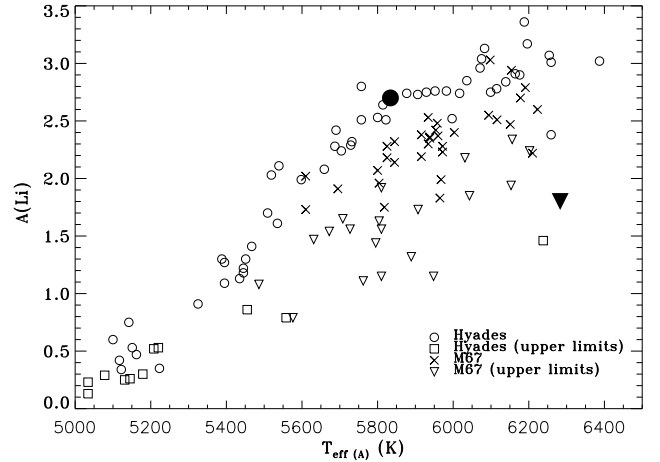}
      \caption{Lithium abundance vs effective temperature for the components of
               \object{HIP 64030} (filled symbols) compared with those
               of stars in the Hyades (Thorburn et al.~\cite{thorburn93}; open circles: 
               stars with Li detection;
               open squares: upper limits) and in M67 (Jones et al.~\cite{jones99}; crosses: stars with Li 
               detection; open triangles: upper limits.}
         \label{f:li_hip64030}
   \end{figure}

The secondary has a more normal abundance pattern. However, the $^{12}$C/$^{13}$C isotopic 
ratio is lower with respect to typical unevolved stars (Gratton et al.~\cite{gratton00}).
The lithium content of the secondary is similar to that
of Hyades star and slightly larger than that of Li-rich stars in the open cluster M67 
of similar temperature (Fig.~\ref{f:li_hip64030}).
This is somewhat unexpected. It suggests
that no significant lithium depletion occurred during the main
sequence lifetime of this star.
Neutron capture elements are normal for both components.
This indicates that the mass transfer event should have 
occurred during the RGB phase of the originally most massive
star or during the AGB phase of a star not massive enough
to produce s-elements enhancements.

\subsubsection{Differential analysis between the components}

\begin{figure}
   \includegraphics[width=9cm]{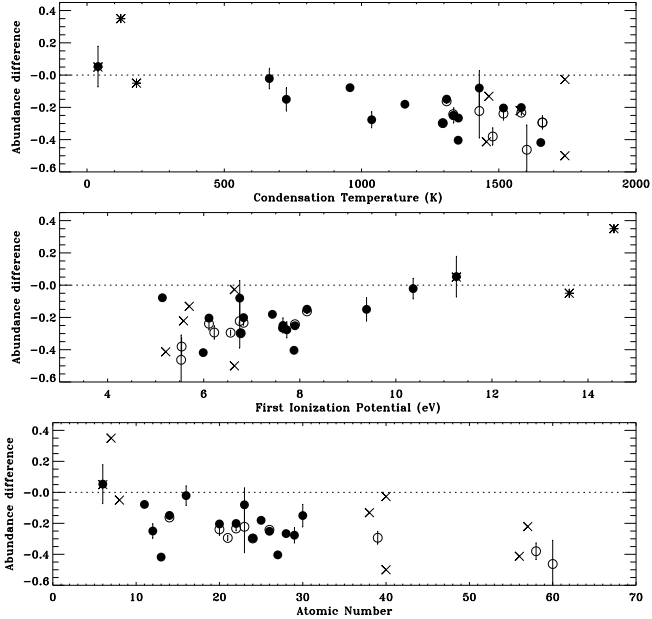}
      \caption{Abundance difference as a function of condensation
       temperature. Filled and empty circles represent abundance
       differences from neutral and single-ionized species
       when at least two lines are available,
       while crosses represent elements for which one single line
       is included in the analysis. Asterisks represent the elements
       for which abundance was obtained by spectral synthesis.}        
         \label{f:trend_abu}
   \end{figure}

The differential analysis provides evidence not only of 
the abundance anomalies caused by the accretion
of CNO processed material by the primary, but also of the large
abundance difference of iron and most of the elements we measured
(Table \ref{t:diffabu}).
Figure \ref{f:trend_abu} shows the abundance difference as 
a function of condensation temperature, first ionization potential,
and atomic number.
A correlation with condensation temperature appears to be present;
the abundance differences also show some correlation  with first
ionization potential, as the latter somewhat correlates with 
condensation temperature.
However, the small abundance difference for sodium indicates
that the real correlation is that with the condensation temperature.

\begin{figure}
   \includegraphics[width=9cm]{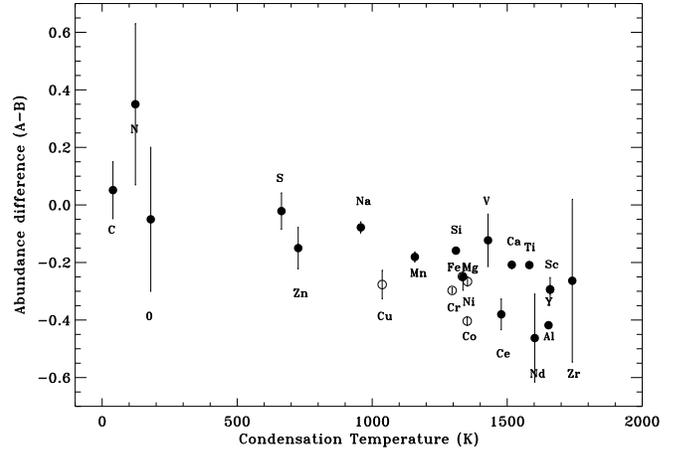}
      \caption{Abundance difference as a function of condensation
       temperature. Only elements with at least two lines are plotted. 
       Abundance difference from neutral and single-ionized species
       were averaged. Empty circles represent siderophiles elements,
       and filled circles the other elements.
       Siderophiles show a larger abundance difference with respect
       to other elements for similar condensation temperatures.}
        
         \label{f:sider}
   \end{figure}

Considering only elements with at least two measured lines,
excluding C and N because of the alterations caused by the transfer
of processed material from the RGB donor,
and averaging the results from neutral and single ionized states
when available (Fig.~\ref{f:sider}), 
the resulting linear correlation coefficient
is -0.63 and the Spearman rank correlation coefficient is
-0.56 with a significance of 98\%.
Therefore, beside the alterations expected by the blue straggler
formation process, the abundance difference between the components
shows the pattern expected in the case of accretion of 
chemically fractionated material: no or low abundance difference for most of 
the elements characterized by a low condensation temperature, and large 
abundance difference for elements with high condensation temperature
(e.g., iron) (Table \ref{t:diffabu}; Fig.~\ref{f:trend_abu}).
 
It must be noted that the condensation temperature of a given element is not
an absolute value, but it depends on the physical conditions and global chemical composition 
of the gas. The adopted condensation temperatures are the '50\% condensation 
temperatures' (those at which half of the atoms of a given element are condensed
and half are in gas phase)
derived by Lodders (\cite{lodders}) for Solar System composition gas.
The $\alpha$ enhancement, other chemical anomalies, and different physical
conditions
may alter the condensation sequence (Lodders \& Fegley \cite{lodders95}).
From Fig.~\ref{f:sider} there is some indication of peculiar
features in the abundance difference vs. condensation temperature
diagram: siderophile elements (Fe, Cu, Cr, Ni, Co) have 
on average a larger
abundance difference with respect to elements with similar
condensation temperature condensing into
oxygen compounds (Mg, Si, V, Ca, Ti). 
The differential analysis does not reveal peculiar differences
of neutron-capture elements, confirming the results of relative
abundances with respect to iron discussed above.


\begin{table*}
\caption[]{Abundance difference between the components of HIP 64030.}
\label{t:diffabu}
\begin{tabular}{cccccccccl}
\hline
\noalign{\smallskip}
Element &  [x/Fe](A)  &  rms &  [x/Fe](B) &  rms  &  $N_{lines}$   & $\Delta$ & rms & err  & Remarks \\
\noalign{\smallskip}
\hline
\noalign{\smallskip}
Fe I     & -0.57$^{1}$&  0.113    & -0.32$^{1}$    &  0.093    & 105    & -0.251    &  0.071    &  0.007 &  \\
Fe II    & -0.57$^{1}$&  0.114    & -0.33$^{1}$    &  0.110    &  35    & -0.242    &  0.062    &  0.010 &  \\
C I      &  0.41    &  0.188    &  0.10    &  0.039    &   3    &  0.053    &  0.217    &  0.125 &  \\
C        &  0.20    &           & -0.10    &           &        &  0.050    &           &  0.140 & synt.\\
N        &  0.50    &           & -0.10    &           &        &  0.350    &           &  0.280 & synt.\\
O I      & $<$0.40  &           &  0.20    &           &   1    & -0.050    &           &  0.250 & [OI] \\
Na I     &  0.14    &  0.004    & -0.03    &  0.031    &   2    & -0.078    &  0.027    &  0.019 &  \\
Mg I     &  0.35    &  0.119    &  0.35    &  0.039    &   3    & -0.250    &  0.081    &  0.047 &  \\
Al I     & -0.38    &  0.091    & -0.21    &  0.079    &   2    & -0.418    &  0.013    &  0.009 &  \\
Si I     &  0.23    &  0.106    &  0.13    &  0.050    &  11    & -0.149    &  0.067    &  0.020 &  \\
Si II    &  0.18    &  0.011    &  0.09    &  0.029    &   2    & -0.162    &  0.018    &  0.013 &  \\
S I      &  0.23    &  0.115    &  0.00    &  0.113    &   3    & -0.021    &  0.109    &  0.063 &  \\
Ca I     &  0.17    &  0.069    &  0.13    &  0.103    &  15    & -0.204    &  0.054    &  0.014 &  \\
Ca II    &  0.16    &  0.129    &  0.15    &  0.183    &   3    & -0.239    &  0.068    &  0.039 &  \\
Sc II    &  0.04    &  0.068    &  0.08    &  0.038    &   8    & -0.294    &  0.042    &  0.015 &  \\
Ti I     & -0.02    &  0.050    & -0.07    &  0.066    &  17    & -0.201    &  0.032    &  0.008 &  \\
Ti II    &  0.04    &  0.076    &  0.02    &  0.079    &  11    & -0.233    &  0.046    &  0.014 &  \\
V I      &  0.18    &  0.280    &  0.02    &  0.139    &  10    & -0.080    &  0.344    &  0.109 &  \\
V II     &  0.21    &  0.500    &  0.18    &  0.735    &   2    & -0.222    &  0.235    &  0.167 &  \\
Cr I     & -0.10    &  0.091    & -0.06    &  0.100    &  29    & -0.297    &  0.065    &  0.012 &  \\
Cr II    & -0.03    &  0.115    &  0.02    &  0.098    &  17    & -0.298    &  0.063    &  0.015 &  \\
Mn I     & -0.20    &  0.009    & -0.27    &  0.029    &   3    & -0.181    &  0.030    &  0.017 &  \\
Co I     & -0.31    &  0.151    & -0.15    &  0.166    &   3    & -0.404    &  0.024    &  0.014 &  \\
Ni I     & -0.09    &  0.103    & -0.07    &  0.059    &  32    & -0.266    &  0.090    &  0.016 &  \\
Cu I     & -0.24    &  0.078    & -0.21    &  0.021    &   3    & -0.277    &  0.086    &  0.049 &  \\
Zn I     & -0.01    &  0.041    & -0.11    &  0.158    &   3    & -0.150    &  0.124    &  0.072 &  \\
Sr I     & -0.24    &           & -0.36    &           &   1    & -0.131    &           &        &  \\
Y II     & -0.18    &  0.060    & -0.14    &  0.121    &   6    & -0.293    &  0.100    &  0.041 &  \\
Zr I     &  0.09    &           &  0.34    &           &   1    & -0.500    &           &        &  \\
Zr II    & -0.23    &           & -0.46    &           &   1    & -0.027    &           &        &  \\
Ba II    &  0.12    &           &  0.29    &           &   1    & -0.413    &           &        &  \\
La II    & -0.10    &           & -0.13    &           &   1    & -0.221    &           &        &  \\
Ce II    &  0.08    &  0.009    &  0.21    &  0.085    &   3    & -0.380    &  0.093    &  0.053 &  \\
Nd II    &  0.22    &  0.172    &  0.43    &  0.341    &   3    & -0.463    &  0.264    &  0.153 &  \\
\noalign{\smallskip}
\hline
\end{tabular}

$^{1}$ [Fe/H]

\end{table*}


\section{Discussion}
\label{s:discussion}

The  conclusion that  can  be drawn  from  the differential  abundance
analysis is that there is a real composition difference between the two
components  of \object{HIP64030},  the primary,  that is  the  blue stragglers,
being more Fe poor by roughly a factor of two.
The uniqueness of the large abundance difference of \object{HIP 64030} among 50
pairs analyzed and its BSS status suggests a link between them.
As discussed in Sect.~\ref{s:intro}, iron is not expected to be altered during
the formation of the blue straggler.
Abundance anomalies should be confined to light elements and, in the case of
an AGB donor, on neutron capture elements.
Therefore, while the signatures expected from mass transfer
from a RGB (or low-mass AGB) star are indeed observed in the blue straggler component
(N enhancement, small $^{12}$C/$^{13}$C isotopic ratio, no lithium), 
other mechanism(s) should play
a fundamental role in the production of the observed 
abundance pattern.

We  will now  discuss two  possible scenarios  of  selective accretion
that may explain the origin of this abundance difference. Both of them envisage a similar
evolution of the system. This  is initially composed of a close binary
(\object{HIP64030Aa}  and   \object{HIP64030Ab})  and  of  a   third  farther  companion
(\object{HIP64030B}). Originally, the most  massive of the stars is \object{HIP64030Ab},
with   a  mass  in   the  range   between  0.9~$M_\odot$\   and  $\sim
1.3-1.5~M_\odot$. The lower limit is  the mass of \object{HIP64030B}, and it is
determined by  the consideration that  its evolution should  be faster
than  that of  \object{HIP64030B},  which is  only  now evolving  off of the  main
sequence.  The  upper limit  is  determined  by  the evolutionary  age
difference between \object{HIP64030Aa} and \object{HIP64030B} ($\sim 5$~Gyr), and by the
consideration  that if  the  mass transfer  occurred  too early,  this
second star would now be  too evolved. The original mass of \object{HIP64030Aa}
is not well known; however, it  should be smaller than the current mass
of 1.04~$M_\odot$.  Hence the total  mass of the system  was originally
lower than $\sim 3.3~M_\odot$.

When \object{HIP64030Ab}  evolved off of the main sequence,  it filled  its Roche
lobe  and  began  transferring  significant amounts  of  material  on
\object{HIP64030Aa}, as well  as losing a considerable amount  of mass (McCrea
channel of  formation of  the blue stragglers).  Notice that  the mass
transfer episode probably occurred during the first  ascent of the
RGB  since there  is no  trace of  triple alpha  reaction  or s-process
products in  the composition of  \object{HIP64030Aa}. 
Alternatively, the mass
transfer episode can also have  occurred during the AGB phase of 
a star not massive enough to produce significant s-element enhancements.
Mass transfer  ended when
the  outer convective  envelope of  \object{HIP64030Ab} was  entirely consumed,
leaving only  the degenerate core, which  is probably a  He-white dwarf of
$\sim   0.2-0.4~M_\odot$.  \object{HIP64030Aa}   acquired  material   from  the
convective envelope of \object{HIP64030Ab}: a  lower limit to the accreted mass
is   given  by   the   mass  difference   with   \object{HIP64030B},  that   is
about 0.1~$M_\odot$.  This is  much more  than  the mass  of its  convective
envelope,  so  that we  may  assume  hereinafter  that its  convective
envelope is entirely composed of material originally in the convective
envelope of  \object{HIP64030Aa}. This agrees well with  its chemical abundance
pattern.

The orbit of the binary   \object{HIP64030A-B} should have widened
considerably during the mass loss episode.
The details of the orbital evolution depend critically on the 
comparison between the timescale
of the mass ejection and the orbital period.
The mass loss by stellar wind before Roche lobe overflow
should have caused a small adiabatic widening of the
orbit ($a_{end}=a_{start}*m_{start}/m_{end}$), without significant
changes to the orbital eccentricity.
For the mass loss during the common envelope phase, the situation
is different, as hydrodynamical simulations 
(e.g., Sandquist et al.~\cite{sandquist98}; Rasio \& Livio \cite{rasio96})
indicate a very rapid ejection ($\sim$ years), much shorter than the
orbital period of the wide component.
In this case, the widening of the orbit can be much larger
(Boersma \cite{boersma}).
In our case, the system remains bound, as the final mass
of the system is $\sim 2.1-2.4~M_\odot$, well above half of  the original mass.
If a major part of mass loss from the central binary 
occurred during the common envelope phase, we may reasonably  guess that
the original separation  between \object{HIP64030A} and B was 
about one half or one third of the present one.
Considering the current projected separation  of 460 AU,
and the orbits compatible with the observations (Sect.~\ref{s:astrometry}), 
the original binary separation may then have been about 100-200 AU.
Closer original separations are  possible if the common envelope
ejection occurred by chance close to the periastron passage of the
wide companion, as in this case the widening of the orbit can be 
much larger (Hills \cite{hills83}).

\subsection{Scenario A. Selective accretion of metal-depleted material on the primary}

The  basic idea of  this scenario  is that  HIP64030B is  a completely
normal star, and its surface  composition is identical to the original
system composition. All anomalies  are then confined to HIP64030Aa. In
this  case there  is a  depletion of  the original  Fe content  at the
surface of HIP64030Aa.

To  explain the anomalous Fe-peak content  of HIP64030Aa, the
idea  is to  call for  a  mechanism similar  to that  found acting  in
post-AGB binaries (Van Winckel \cite{vanwinckel03}). 
Large  Fe depletions (by several orders of
magnitude) have been observed in these stars. The mechanism called for
envisages the formation of a circumbinary torus, from which dust is
selectively removed by radiation  pressure and high velocity wind. Part
of the remaining (dust-free) gas is then accreted again by the central
star (Waters et al.~\cite{waters92}). 
In the case of post-AGB stars, the amount of gas accreted by the
star  need  not  be  large because  there  is  not any  convective
envelope. In the case of  HIP64030Aa, this should be comparable to the
mass ($\leq 0.01~M_\odot$) of  the outer convective envelope, to cause
enough dilution effect for Fe.

\begin{figure}
   \includegraphics[width=9cm]{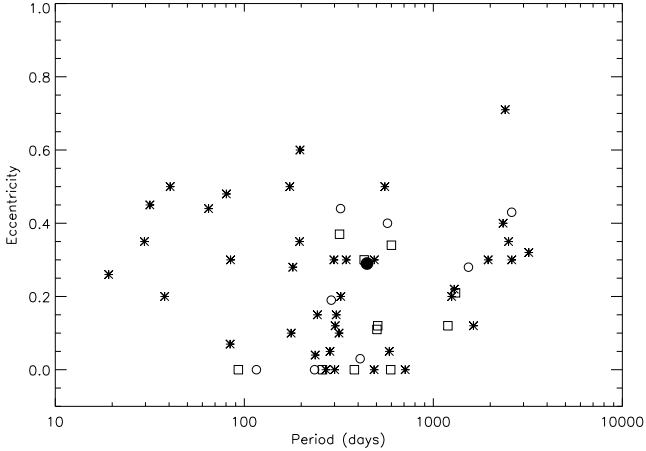}
      \caption{Period eccentricity relation for post AGB and RV Tauri
               stars with depleted abundances (empty squares)
               and normal abundances (empty circles) from Maas (\cite{maas03})
               and of field blue stragglers studied by Preston \& Sneden (\cite{preston00})
               and Carney et al.~(\cite{carney05}) (asterisks). The position of
               HIP 64030 is overplotted as a filled circle.}
        
         \label{f:postagb}
   \end{figure}

The RV Tauri and post AGB with selective depletion are
mostly and likely all binaries (Van Winckel et al.~\cite{vanwinckel99}).
Their orbital periods and eccentricity overlap
with those of field blue stragglers (Fig.~\ref{f:postagb}).
The accretion of  chemically
fractionated material observed on the post-AGB component
should also occur  on the companion.
Such a companion, if the amount material received during
the mass transfer event is large enough, will become a blue straggler, while the ex-post AGB
component will become a faint white dwarf. A possible case of a future blue 
straggler among the companion of  post-AGB binaries
is represented by the companion of \object{IRAS 05208-2035} (Maas \cite{maas03}).

While Keplerian disks appear to be common around post-AGB binaries
(De Ruyter et al.~\cite{deruyter06}),
it is not clear if a  significant fraction of the
heavy material may really condense onto dust in a torus around an RGB
star (probably the case of HIP64030A).
If this mechanism  is really active for \object{HIP 64030},
it  should  be  so in  other  blue  stragglers  formed with  the  McCrea
mechanism. 
However, we are not aware of other evidence for chemically fractionated accretion
on blue stragglers. 
Blue stragglers  observed in \object{M67} (Shetrone \& Sandquist \cite{shetrone}) share
the same Fe  content of the other cluster star (one case with lower metal
abundance is probably due to the composite nature of the spectrum).
For field blue stragglers, it is not easy to identify anomalies in 
element abundances and abundance ratios at the level of 0.2 dex.
Anyway, available literature data for field blue stragglers are quite sparse.
Most of abundance analysis of these stars do not
include key elements to test the accretion 
of chemically fractionated material (e.g., sodium, sulfur, zinc),
focusing their attention mostly on the search for anomalies
of light elements and neutron capture elements (e.g., Sneden et al.~\cite{sneden03}).
It should also be considered that
the gas-dust separation might be at work only above a metallicity threshold
(e.g., [Fe/H]=--1.0, as suggested by Giridhar et al.~\cite{giridhar00} 
for RV Tauri stars). If this
is the case, the pattern should not be observed in the most metal poor
blue stragglers.
Finally, this mechanism  does not explain
the  unusual $^{12}$C/$^{13}$C isotopic  ratio found
for  HIP64030B in a  natural way.   We  are  then   forced  to  suggest   an  independent
(primordial) origin for it (the high lithium content of the secondary
allows us to rule out the onset of first dredge-up for this star).

\subsection{Scenario B. Selective accretion of metal-rich material on the secondary}

The  basic idea  of  this scenario  is  that HIP64030Aa  has a  normal
abundance pattern for its  evolutionary status, and that the abundance
difference  found between  the  two  visible components  is  due to  a
combination of the classical evolutionary pattern of a blue straggler,
and of selective accretion on  HIP64030B. In this scheme, the original
Fe-peak element abundances of the system is that found in HIP64030Aa.

There are two basic assumptions in this scenario:
\begin{itemize}
\item  HIP64030B should  have  been  able to  capture  a large  enough
fraction of  the material expelled from  the primary (which  is of the
order  of 1~$M_\odot$).  This should  be  at least  comparable to  the
present  mass of  its convective  envelope  ($\sim 0.015-0.030~M_\odot$)
because it should  include an amount of Fe  large enough to roughly
double  its  original Fe  abundance.  
\item There  should have been some selective  accretion mechanism. The
basic idea  here is  that the accreting  material might have  formed a
disk around \object{HIP64030B},  similar to a proto-planetary disk.  On a short
timescale, dust  settled  at the equator  of this disk.  Then, the
same mechanism  called for planet migration may have been onset, causing infall
of  rocky  material on  the  central  star.  Since when  this  episode
occurred  \object{HIP64030B}  was  already   a  well  evolved,  old  star,  the
convective envelope  was already in  place when this disk  evolved, so
that quite small amounts of  rocky material (a few Earth masses) would
be enough to  significantly enrich the surface of the star.
\end{itemize} 

To evaluate  the plausibility of the occurrence of an adequate 
amount of accretion on \object{HIP 64030B} (about $0.01~M_{\odot}$), 
we first considered
the prescriptions of the Bondi-Hoyle wind accretion 
(see, e.g., Hurley et al.~\cite{hurley02}).
It results that for  original masses of the inner pair 
$\sim1.5~M{\odot}$ and $\sim0.9~M{\odot}$, 
a binary separation as small as 20-30 AU is required, while
at 100 AU the accreted mass should be of the order of  $0.001~M_{\odot}$.
Hydrodynamical simulations by Theuns et al.~(\cite{theuns96}) found
 lower accretion rates with respect to the standard  Bondi-Hoyle
ones, when the wind speed is comparable to orbital velocity, 
which occurs for separation smaller than about 20 AU.
It must be noted that these simulations refer to the
accretion of mass lost by a single star through stellar wind, while
we expect that the blue straggler \object{HIP 64030A} 
evolved through Roche lobe overflow and common envelope phase.
The velocity, time and spatial distribution, and other properties
of the material lost by the inner binary should be different
with respect to the stellar wind from an isolated star,
possibly significantly affecting  the amount of accreted
material by a wide companion.
The available models of the ejection of the common envelope 
show large spatial inhomogeneities in the outflow, with most
of the material ejected in the orbital plane of the close
binary (see, e.g., Sandquist et al.~\cite{sandquist98}). Therefore,
a suitable orientation of the orbital planes of the triple
system might provide larger accretion rates, if the ejection
velocity is similar to that of stellar wind. 
Further studies are required to evaluate the plausibility of
this scenario.

As  possible  pros of this scenario, we notice  that with respect to
Scenario  A,  in  this  scenario  we may  invoke  the  same  selective
accretion mechanism to  explain the unusual $^{12}$C/$^{13}$C isotopic
ratio found  for HIP64030B. On the  other hand, it  might require some
fine tuning in order for enough matter to be accreted by \object{HIP64030B}.
The rather high lithium content of \object{HIP64030B} is not naturally
explained by this scenario, as the material falling into the star
should have been lithium-depleted.
The validity of this scenario can be tested by looking
for abundance anomalies of wide companions orbiting 
blue stragglers, other stars on which mass transfer from
an evolved companion occurred (e.g., barium stars, CH stars), and
central stars of planetary nebulae.


\section{Summary}
\label{s:conclusion}

The system \object{HIP 64030} is composed of two moderately metal-poor 
solar-mass stars with a projected separation of about 500 AU, with
the primary being a spectroscopic binary with a period of 445 days
and a companion mass larger than $0.17~M_{\odot}$.
The (currently) most massive component appears to be a blue
straggler.
High-precision differential abundance analysis reveals
the signatures of the mass transfer expected for a blue straggler
formed via McCrea mechanism, but also an unexpected difference
of most of the other elements, including iron, correlated
with the condensation temperature.

Two scenarios were devised to explain the observed abundance pattern.
In the first one, all the abundance anomalies occur on the blue
straggler. Its lower metal abundance would be explained by the
accretion of dust-depleted material, in a similar way to what
has occurred for several post-AGB binaries.
The second scenario is based on the accretion of dust-rich
material on the secondary.
It may at least qualitatively explain the whole
abundance pattern (including the low $^{12}$C/$^{13}$C of the secondary,
unexpected for a normal main sequence star). 
However, it requires that a rather large amount of
material lost by the blue straggler system be captured by
the secondary. 
Whether this is actually realistic would require a dedicated
modeling.
A refinement of the constraints available for the binary orbit
(new high-precision relative astrometric observations,
improvement on distance estimate) would also make
the evaluation of this scenario easier.

The reality of either scenario implies testable predictions.
For the first scenario, 
alterations of iron abundance and trends of abundance with
condensation temperature should be observed
in other blue straggler formed via mass transfer.
For the second scenario, signature of accretion of chemically 
processed material should be found for wide companions
of blue stragglers and other related objects.

\begin{acknowledgements}

   This research has made use of the 
   SIMBAD database, operated at CDS, Strasbourg, France. 
   We warmly thank R. Cosentino, A. Martinez Fiorenzano, R. Wittenmyer, and W. Cochran 
   for the observations at TNG and Mc Donald Observatory.
   We thank M. Barbieri for the computation of the galactic orbit.
   We ackowledge throughful comments by the referee Dr. Preston. 
   This work was partially funded by COFIN 2004 
   ``From stars to planets: accretion, disk evolution, and
   planet formation'' by Ministero Universit\`a  e Ricerca
   Scientifica Italy. M.~Endl is supported by the 
   National Aeronautics and Space Administration
   under grants NNG 04-G141G and NNG 05-G107G.

\end{acknowledgements}

\end{document}